\documentclass[10pt,journal,final]{IEEEtran}

\usepackage{soul}

\ifCLASSOPTIONcompsoc
  \usepackage[nocompress]{cite}
\else
  \usepackage{cite}
\fi
\ifCLASSINFOpdf
\else
\fi
\usepackage{graphicx}
\usepackage{subfigure}
\usepackage{epstopdf}
\usepackage{tcolorbox}
\graphicspath{{figure/}}

\usepackage{float}

\usepackage{algorithm}
\usepackage{algorithmic}

\usepackage{booktabs}

\usepackage{color}

\begin{document}
%
\title{Sitting Posture Recognition \\ Using a Spiking Neural Network}
%
%
%
%

\author{Jianquan~Wang, Basim Hafidh,
Haiwei~Dong,~\IEEEmembership{Senior~Member,~IEEE}
and~Abdulmotaleb~El~Saddik,~\IEEEmembership{Fellow,~IEEE}
\IEEEcompsocitemizethanks{\IEEEcompsocthanksitem The authors are with the Multimedia Computing Research Laboratory, School of Electrical Engineering and Computer Science, University of
Ottawa, Ottawa, ON K1N 6N5, Canada (e-mail: jwang438@uottawa.ca; basim.hafidh@uottawa.ca;
haiwei.dong@ieee.org; elsaddik@uottawa.ca).\protect\\
}
\thanks{}}

%
%

\markboth{IEEE SENSORS JOURNAL}%
{Shell \MakeLowercase{\textit{et al.}}: Bare Advanced Demo of IEEEtran.cls for IEEE Computer Society Journals}
%



\IEEEtitleabstractindextext{%
\begin{abstract}
To increase the quality of citizens' lives, we designed a personalized smart chair system to recognize sitting behaviors. The system can receive surface pressure data from the designed sensor and provide feedback for guiding the user towards proper sitting postures. We used a liquid state machine and a logistic regression classifier to construct a spiking neural network for classifying 15 sitting postures. To allow this system to read our pressure data into the spiking neurons, we designed an algorithm to encode map-like data into cosine-rank sparsity data. The experimental results consisting of 15 sitting postures from 19 participants show that the prediction precision of our SNN is 88.52\%.
\end{abstract}

}

\maketitle

\IEEEdisplaynontitleabstractindextext

%
\IEEEpeerreviewmaketitle
\ifCLASSOPTIONcompsoc
\IEEEraisesectionheading{\section{Introduction}\label{sec:introduction}}
\else
\section{Introduction}
\label{sec:introduction}
\fi

In recent years, many studies have been conducted to improve the quality of life of citizens in smart cities. One of the visions towards smart cities is digital twins \cite{el2018digital}, which are a replica of any living or nonliving entity. According to a Global Burden of Disease (GBD) study \cite{vos2017global}, increasingly more people are suffering from lower back pain among other conditions due to inappropriate sitting behaviors. To improve quality of life, it is essential to design personalized sensing systems that learn the sitting behavior or monitor the health condition of individuals and provide real-time multimodal feedback (through sound, video, haptics, etc.) \cite{hua2019monitoring}. The work presented in this paper is an attempt towards deep learning-powered chairs that a) teach individuals subjects the best sitting posture out of 15 postures and b) provide real-time feedback in case the subject is not sitting properly. We designed two arrays of pressure sensors to obtain the pressure data of different sitting postures so that we can make a ``smart'' chair to identify whether the posture is good or bad for the well-being of the subject. To obtain highly credible results, we should eliminate any intervention behavior during the training phase when the system is establishing the dataset and classifying the subjects' sitting postures. Thus, a controlled calibration measurement procedure is required. The generated data are correctly and automatically labeled and used for training a deep learning machine.

The backpropagation neural network in machine learning uses the McCulloch-Pitts (M-P) neuron model, which reflects one or more important neurophysiological observations, without regard to other neurophysiological facts. Although amenable to mathematical analysis, such an empirical model has no rigorous connection with real neural systems \cite{hoppensteadt2012weakly}. The neuron model in a spiking neural network, such as a Hodgkin-Huxley model, reflecting more neurophysiological observations than the 
{M-P} 
model, is still incomplete. However, additional mechanisms provide advantages. For example, the time interval between each spike implies that the neuron carries temporal information. Therefore, we decided to try a more complex neuron model to challenge the machine learning task.

Liquid state machines (LSMs) are one of the most effective computation modules   and were proposed by Maass \cite{maass2002real}. They use a recurrent network structure so that the intermediate state of the network is related not only to the input at the present time but also to that at the previous moment. That is, the membrane potential of the neurons is related to the quantity, frequency and interval of the input spikes and not just one spike. In this paper, a spiking neural network is constructed in the form of a liquid state machine. The purpose of this work was to design, implement, and validate a sensing chair system for computer-human interactions with the spiking neural network.

This paper is organized as follows: In Section \ref{sec:Related Work}, an overview of existing related technologies and research work is presented. The seat system is described in Section \ref{sec:System Description and Pressure Sensor}. Section \ref{sec:Posture Recognition Approach} provides the details of the recognition model and algorithms. Section \ref{sec:Experiment Results} provides the experimental details for testing the models and the results of the classification. Section \ref{sec:Conclusions} concludes the paper.

\section{Related Work}
\label{sec:Related Work}
\subsection{Pressure Sensing}
\begin{figure*}[ht]
\centering
\includegraphics[width=7in]{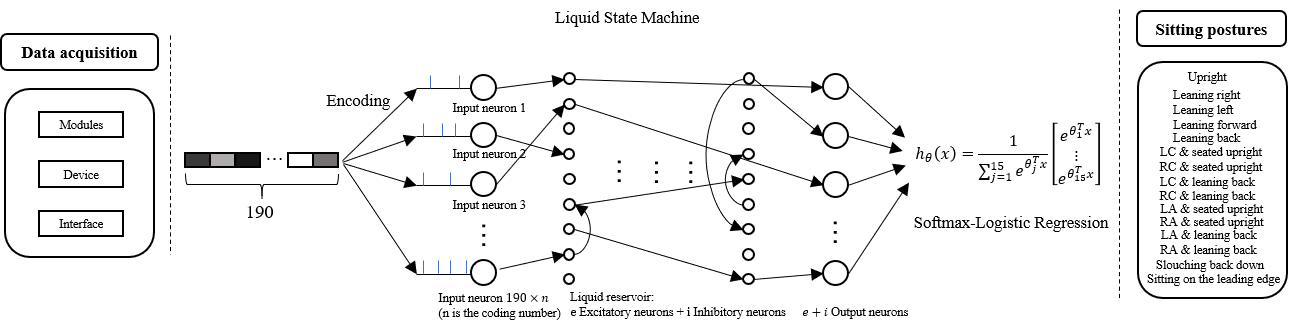}
\caption{The structure of the proposed system. The data acquisition module is designed based on modules with different functions, hardware devices and software interfaces, which are introduced in Section \ref{sec:System Description and Pressure Sensor}. In the median part, the data processing module shows the overall machine learning process, which is introduced in Section \ref{sec:Posture Recognition Approach}. The sitting posture module shows a total of 15 sitting postures in detail that we aim to recognize. ``LC" stands for ``left leg crossed over the right leg". ``RC" stands for ``right leg crossed over the left leg". ``LA" stands for ``left ankle resting on the leg". ``RA" stands for ``right ankle resting on the leg".}
\label{fig_sim}
\end{figure*}
In recent years, weight-supporting surfaces in shoes, beds, and chairs have been widely analyzed using pressure-mapping sensors, which are much cheaper than computer vision systems \cite{Basim_2017}. Examples of shoe studies include assessing pressure discomfort thresholds, assessing pressure pain thresholds and tissue stiffness on the plantar surface of the foot\cite{weerasinghe2017pressure}, formulating biomechanical foot models to assess strains within the foot and to determine the risk of ulcer formation\cite{bucki2016clinical}, and studying the mechanics of a healthy gait during gentle turns\cite{peyer2017locomotion}. Examples of bed studies include helping caregivers adjust bedridden patients to the correct sleeping posture to reduce the risk of pressure ulcers \cite{hsia2009analysis, yousefi2011bed} and developing a monitoring framework for improving patient safety and caregiving efficacy\cite{hung2015bed}. Examples of chair studies include monitoring sitting behavior for healthcare services, human-computer interactions, and intelligent environments\cite{huang2017smart}, developing a driving assistant for reducing traffic accidents by estimating the driver's sitting posture and warning the driver\cite{ding2017estimation}, using eight pressure sensors as a part of a posture determination system to give patients feedback on whether their sitting posture is correct\cite{ahn2015implemented}. Most of the previous experiments, such as the work in \cite{tan2001sensing,mota2003automated,zhu2003template}, use a sensing chair equipped with a commercial pressure distribution sensor sheet called the body pressure measurement system (BPMS) manufactured by Tekscan. However, this sheet is very expensive. Therefore, we designed our sensing chair using force-sensitive resistors (FSRs) that were combined to form a sensing grid (see Fig \ref{subfig:grid}). The FSR is a variable resistor that changes based on the pressure applied to its area. Our pressure-mapped sensor on the chair is introduced in the next section.

\subsection{Smart Chair}
The smart chair is designed to process real-time information for the user. 
{It includes chairs for different purposes, sensors, microcontrollers, and classification ability.} 
\cite{lee2015feasibility} gave a feasibility study that showed the pressure matrix is possible to provide data in the sitting posture classification. A smart chair was designed in \cite{huang2017smart} for monitoring the sitting behavior and was composed of an 8 by 8 pressure sensor array, data acquisition module, and a computational terminal, which could recognize 8 postures. The chair had only 1 sensor array that laid on the seat, whereas ours have 2 separate sensor arrays (on the seat and on the back of the chair), which supports the classification of 15 postures. Ding et al. \cite{ding2017estimation} developed a smart chair with pressure sensor sheets set up on the seat and back of the driver's seat for estimating driving postures. The sensor sheets had 16x16x2 sensor units. However, the system could only recognize 8 postures. The driver's smart chair did not include any feedback mechanism to alert the driver to a risky posture. A smart chair with a feedback mechanism was proposed in \cite{ahn2015implemented}, which had sensor cushion on the seat. It was soft and comfortable for the patient but was too thick to be flexibly installed on different types of chairs. The feedback signal could be directly shown in the smartphone app. In contrast to the feedback information, our chair has two vibration motors on the armrest, which are sufficiently flexible to be installed on other places, such as the back waist of the user. Such haptic feedback is more responsive than the information display.
\subsection{Spiking Neural Network}

Typically, the second generation of artificial neural network (ANNs) developed in the 1980s is considered a simulation of biological neural networks. Nowadays, ANN has been successfully used in classification tasks \cite{cao2018convolutional}. However, there is ample evidence that timing phenomena such as time differences between spikes and the frequency of the oscillating subsystem are part of various information-processing mechanisms in biological nervous systems \cite{aertsen1993brain}. Thus, the formal model of the 3rd generation of ANNs called spiking neural network (SNN)  for modeling such timing phenomena of neural networks was defined by Maass \cite{maass1995computational}.

With the development of SNNs, learning algorithms based on SNNs have also emerged. The earliest emergence of unsupervised learning algorithms, usually based on Hebb's rule, was followed by more biologically 
spike-timing-dependent-plasticity (STDP)
learning rules, which have become a research hotspot \cite{hebb2005organizations}. The traditional artificial neural network supervised learning algorithm refers to adjusting the weight of the neural network according to the obtained error relative to the actual output value of the neural network compared with the target value. Related experiments show that there are supervised learning strategies in biological neural networks, especially in motion-aware networks\cite{knudsen1994supervised}, but it is still unclear how biological neural networks implement this mechanism. SNNs are the closest models to biological neural networks. A spike time series represents the information it conveys. The state and error function inside the neuron do not satisfy the conditions of continuousness and differentiability. Therefore, a traditional neural network supervised learning algorithm, such as the BP algorithm, cannot be directly used. According to the difference in the weight adjustment rules, the supervised learning algorithms for SNNs can be approximately divided into gradient descent learning algorithms, supervised STDP learning algorithms, and learning algorithms based on spike sequence convolution\cite{kasinski2006comparison}.

\section{System Description and Pressure Sensor}
\label{sec:System Description and Pressure Sensor}

\subsection{System Description}
The illustration of the working stream in our proposed system is shown in Fig \ref{fig_sim}. We designed the hardware instrument and software interface for data acquisition. The collected data can be processed in real-time or stored for later. In total, 15 common sitting postures were designed to be recognized by our system. We trained our network by using precollected data to be able to identify real-time sitting postures.
\subsection{Hardware Description}
Our proposed chair is composed of a transducer module, a data collection module, and a calibration module. The transducer module represents a set of interconnected sensors and actuator units \cite{Basim_2015}. Each sensor unit is composed of a force-sensitive resistor (FSR), a multiplexer and an analog-to-digital converter (ADC). Two FSR sheets were designed for the seat pan (9x9) and backrest (10x9) of the chair. The FSR unit converts the pressure applied to each unit into an 
{analog} 
voltage. The output of each FSR unit is transformed into a 10-bit pressure reading. The actuator unit is activated to run the assigned actuator(s) upon receiving control signals from the data collection module. This module provides the necessary space to collect all the measured data that comes from the transducer module and passes them to the interface module. It sends control signals to the multiplexers in the sensor module for both the seat pan and backrest. It also receives control signals from the posture classifier and passes them to the assigned haptic actuators. The calibration module is responsible for calibrating the transducer units before each test to improve the accuracy of the pressure measurements for each FSR unit.

The microcontroller used is an Arduino Mega 2560, an open source single-chip microcontroller that has an 8-bit ATmega 328 processor running at a 16-MHz clock speed. The Arduino software runs the data collection and calibration modules previously described. Since the number of analog input ports (ADCs) for the microcontroller is limited, 12 16-bit analog multiplexers were used. The microcontroller also receives control signals from the posture recognizer and passes them to the correct actuator when the sitting posture is incorrect.
\begin{figure}[ht]
\subfigure[Our pressure sensing grid. ]{
\includegraphics[width=1in]{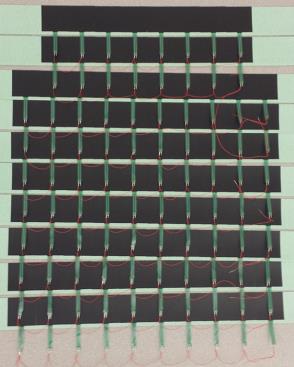}
\label{subfig:grid}
}
\centering
\subfigure[The front view of the chair. ]{
\includegraphics[width=1in]{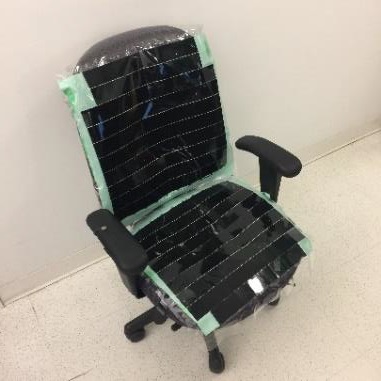}
\label{subfig:front}
}
\subfigure[The back view of the chair. ]{
\includegraphics[width=1in]{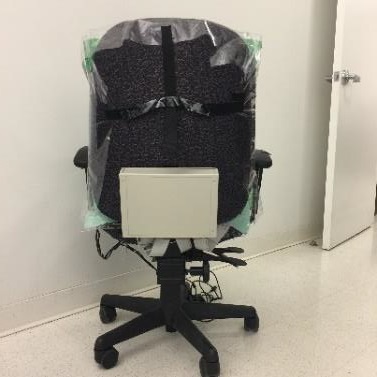}
\label{subfig:back}
}
\subfigure[The user interface. ]{
\includegraphics[width=3in]{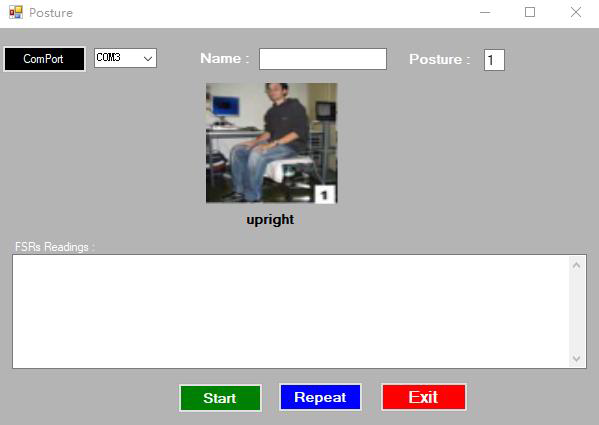}
\label{subfig:interface}
}
\caption{View of the sensing grid and the seat system. The sensing grid is bound to the chair and connected to the control unit, which is inside the box and attached to the back of the chair. The user interface is used for data acquisition and shows the classification result.}
\label{fig:seat system}
\end{figure}
\subsection{Software Description}
The user interface (see Fig \ref{subfig:interface}) provides the necessary communication between the chair and its external environment. Three possible communication media can be used with the chair, including USB (RS232), Bluetooth technology (IEEE 802.15.1), and ZigBee (IEEE 802.15.4). Our proposed system supports both Bluetooth and USB. This module is responsible for exchanging the data and control signals between the data collection module and the posture classifier. The posture classifier module receives the sensor data from the data collection module, stores these data in a database, analyzes them, and informs the user whether his/her posture is correct. The module uses the machine learning mechanism to investigate the poses and sends signals to activate embedded actuators if the user's posture is incorrect. This module is explained in detail in the next section.

\section{Posture Recognition Approach}
\label{sec:Posture Recognition Approach}
\subsection{Spiking Neuron and Synapse}
\begin{figure}[hb]
\centering
\includegraphics[width=3in]{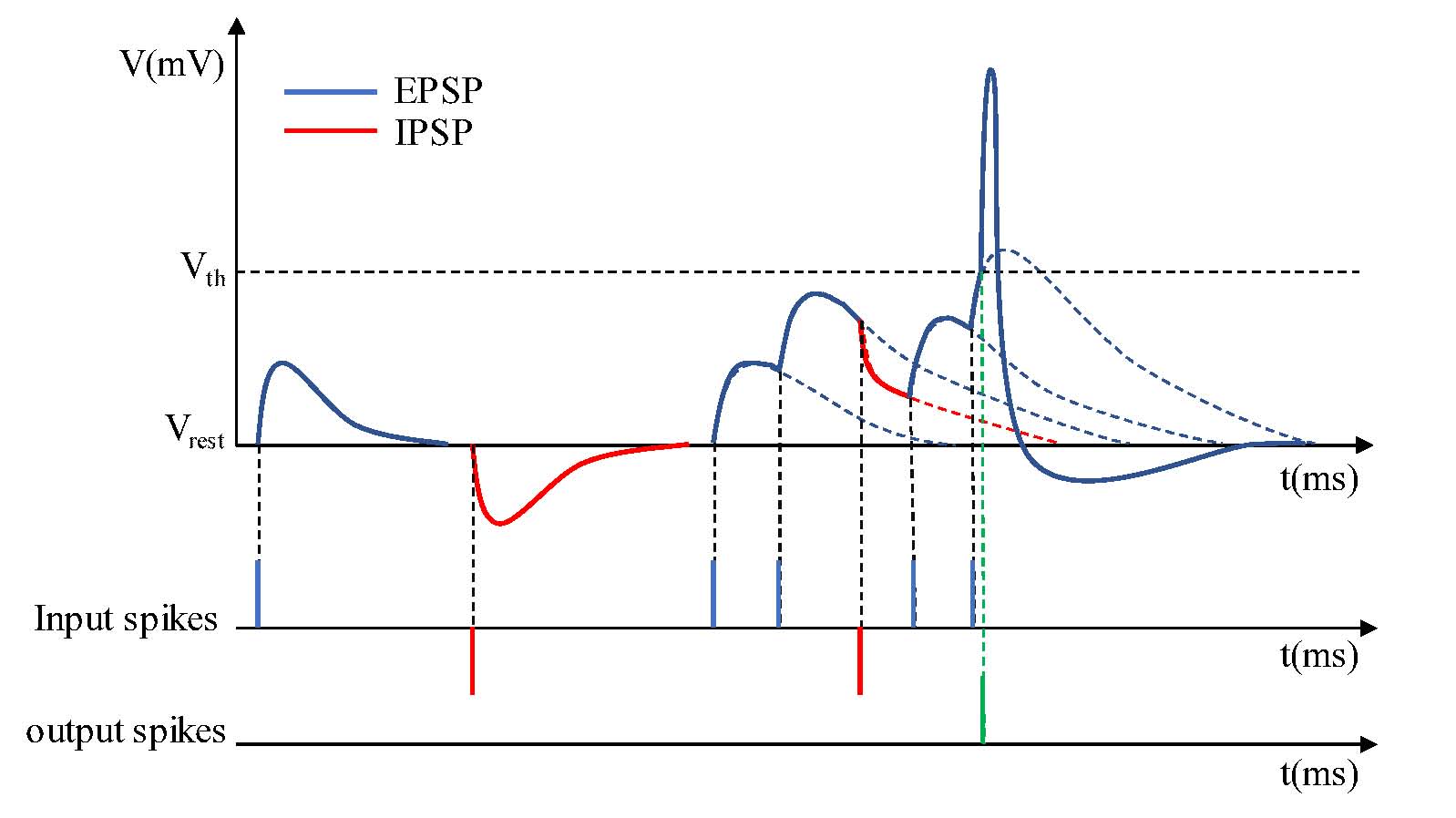}
\caption{The firing progress of the spiking neuron. }
\label{fig:fire}
\end{figure}

The condition of the neurons in the SNN is determined by the membrane potential and the activation threshold. The membrane potential in the neuron is the additive effect of the post-synapse potential connected to the neuron. The post-synapse potential can be divided into the excitatory post-synapse potential (EPSP) and the inhibitory post-synapse potential (IPSP). The membrane potential will increase when receiving the EPSP and decrease when receiving the IPSP. The neuron fires (spikes) when the membrane potential reaches the activation threshold and drops to the resting potential until the refractory period ends \cite{muzy2017iterative}. During the refractory period, the membrane potential does not change regardless of how the post-synaptic potential changes. The firing progress is shown in Fig \ref{fig:fire}.

The neuron model we used is called the leaky integrate-and-fire (LIF) model\cite{burkitt2006review}, which i) is more biologically realistic than the generalized IF neuron model and ii) has less complexity than Hodgkin-Huxley (HH) models \cite{hodgkin1952quantitative}. The LIF neuron is modeled as follows:
\begin{equation}
    \tau_{m}\frac{\mathrm{d} \upsilon }{\mathrm{d} t} = -\upsilon (t) + RI(t)
\end{equation}
where $\tau_{m}$ represents the membrane time constant, $\upsilon(t)$ is the membrane potential at time t, R is the membrane resistance, and $I(t)$ is the input of the neuron. In this leaky integrator equation, the initial $\upsilon$ is the rest potential $\upsilon_{rest}$.
Similar to an ANN's synapse, the weight assigned to each synapse affects the membrane potential of the target neuron. When a spike occurs in the presynaptic neuron, it causes an instantaneous change to occur in the postsynaptic neuron.

\subsection{Liquid State Machine}
\begin{figure}[ht]
\centering
    \includegraphics[width=3.5in]{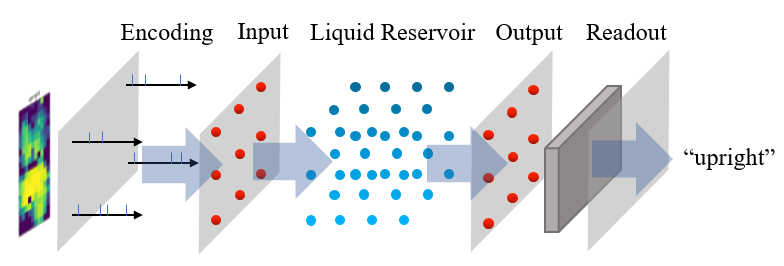}
\caption{Construction of LSM model. The input neurons randomly connect with excitatory neurons and inhibitory neurons. Excitatory neurons can connect to other excitatory neurons and inhibitory neurons by probability. For each excitatory and inhibitory neuron, there is an output neuron connected to it. The number of neurons we used is larger than that shown in this figure.}
\label{fig:SNNmodel}
\end{figure}
A liquid state machine (LSM) is a type of recurrent neural network that is suitable for dealing with dynamic problems and has two macroscopic properties: \textbf{a)} the separation property addresses the difference in liquid patterns from two different input stimuli, and \textbf{b)} the approximation property addresses the resolution and recording capabilities of the readout mechanisms.

As seen in Fig \ref{fig:SNNmodel}, the LSM is composed of an input layer, liquid reservoir and readout layer. The input layer is a set of input neurons for receiving the external stimuli and transferring them into a spike sequence based on neural computation. The liquid reservoir $L^M$ is a liquid filter that maps the signal $u(\cdot)$ from the input layer onto liquid state $X^M(t)$:
\begin{equation}
    x^M\left(t\right)=L^Mu(t)
\end{equation}

In contrast to a finite state machine, the internal structure of the LSM does not need to be designed for a specific task. The topology of the connection between neurons is recurrent and dynamic. The connection and the response of the same perturbation in the reservoir are unequal and unique every time the network is run. The readout layer is the detector, which can be trained to recognize patterns $X^M$ from the liquid filter. The detector can be an SVM, perceptron, logistic regression (LR), or $softmax$ classifier. Such detectors can be seen as a function that maps the liquid state to the target region $y(t)$:
\begin{equation}
    y\left(t\right)=f^M\left(x^M\left(t\right)\right)
\end{equation}

In this article, the reservoir of the LSM comprises $75\%$ excitatory and $25\%$ inhibitory neurons. 30\% excitatory neurons and inhibitory neurons are chosen to be fully connected with the input neurons, followed by randomly dropping out 99\% synapses among them. The mathematical expectation of the number of reservoir neurons per input neuron is 60 if we consider the parameters mentioned in Section V-B. The excitatory and inhibitory neuron group connected and inner-grouped by different spatial distances or probabilities is as follows:
\begin{equation}
    P=C\cdot e^{(-D^2(a,b)/\lambda^2)}
\label{eq:prob}
\end{equation}
where $D(a,b)$ is the Euclidean distance between neuron $a$ and neuron $b$. $\lambda$ is a parameter for adjusting the distance between each connected neuron pair to control the number of synapses. $C$ is specific for different synapses between neuron $a$ and neuron $b$ depending on whether the neurons are excitatory $(E)$ or inhibitory $(I)$. $EE$ in this paper stands for the synapse between two excitatory neurons, $EI$ stands for the synapse between an excitatory and an inhibitory neuron, $IE$ stands for the synapse between an inhibitory and an excitatory neuron (different from $EI$), and $II$ stands for the synapse between two inhibitory neurons.

\subsection{Encoding Method}
\begin{figure}[ht]
    \centering
\subfigure{
    \includegraphics[width=3.5in]{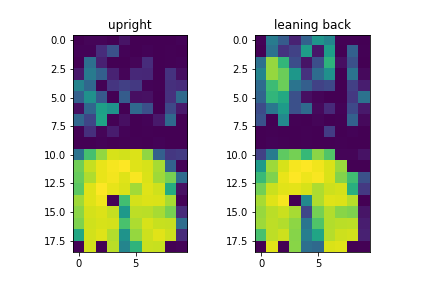}
}
\subfigure{
    \includegraphics[width=3.5in]{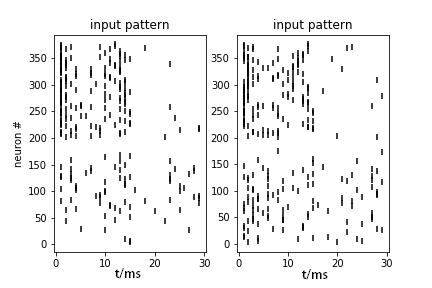}
}
\subfigure{
    \includegraphics[width=3.5in]{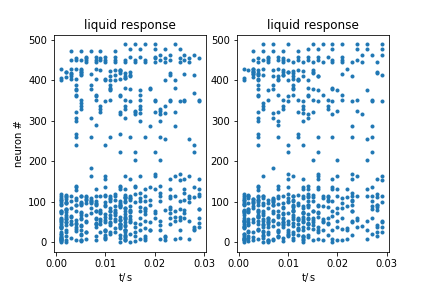}
}
\caption{Converting pressure heat-maps to liquid states. The first row shows the pressure matrix data before encoding. The second row shows the encoded 0/1 matrix data before sending it into LSM, the vertical line in the figure means 1 in the matrix, whereas the blank area means 0 in the matrix. The third row shows the output pattern of LSM, which is called the liquid response.}
    \label{fig:app}
\end{figure}

\begin{algorithm}
\caption{Algorithm for Encoding}
\begin{algorithmic}[1]
 \renewcommand{\algorithmicrequire}{\textbf{Input:}}
 \renewcommand{\algorithmicensure}{\textbf{Output:}}
\REQUIRE {Pressure matrix data $X$ with shape $(p, q)$, Amplitude $A$, Coding number $n$}
 \ENSURE  Cosine-ranked encoded data $E$ with shape $(p\cdot q \cdot n, A)$
\\ \textit{Initialization} : Set $E$ as a zero matrix
  \STATE map $X$ to $[0, \pi]$
\\ \textit{LOOP Process}
\FOR {$i = 0$ to $n-1$}
  \STATE map $X->Y$ with $f(x)=0.5*A*(cos(x+\pi*i/n)+1)$
  \STATE clip $Y$ to $[0, A-1]$
\FOR {each element $y_{j,k}$ in $Y$}
  \IF {($y_{j,k} \ne 0$)}
  \STATE $E_{y_{j,k}+A*j,n*k+i}=1$
  \ENDIF
\ENDFOR
\ENDFOR
 \RETURN $E$
\end{algorithmic}
 \label{algorithm}
\end{algorithm}
The pressure matrix cannot be directly fed into the network since the neuron can only compute a time-varying stimuli train. The values on the pressure map only have spatial relationships with each other. For encoding the pressure matrix to timed stimuli sequences, we designed the algorithm for mapping the matrix data to a cosine-ranked 0/1 matrix with sparsity.
The details of the algorithm are shown in Algorithm \ref{algorithm}. We encoded the data $X$ with shape $(p, q)$ into a 0/1 matrix $E$ with shape $(p\cdot q \cdot n, A)$. The scalar ratios $A$ and $n$ were designed to adjust the sparsity of $E$. $X$ was normalized to $[0, \pi]$ so that the cosine mapping $Y$ was monotonic in the range of $[0, A]$. Note that $Y$ has the same shape as $X$. The value of $Y$ and the corresponding index jointly determine the value in $E$.
Two application examples are given in Fig \ref{fig:app}, which shows the converting process from pressure maps to liquid states. The first row contains two visualized pressure maps as input data, with the shape of 19x10. The left pressure map is labeled as upright posture, and the right one is labeled as leaning back. It was difficult to recognize sitting upright and leaning back for the participant with kyphosis. The second row in Fig 5 contains the visualization of the encoded data. The coding number n is 2 and the amplitude A is 30, therefore the shape of the encoded data is 380x30. Each black vertical line represents 1 in the matrix, and the blank area is filled by 0. With encoding and processing by the LSM, the liquid states showed the spatiotemporal pattern of liquid displacement. Each point in the liquid response means that the neuron is fired at that time.

\section{Experiment and Results}
\label{sec:Experiment Results}
\subsection{Data Acquisition}
We collected data from 19 participants: {14 males and 5 females. The age of participants ranges from 22 to 58, with a median of 25 and a mean of 28.53. The maximum pressure values (representing the weight info) for the participants range from 691 to 1023, with a median of 1007 and an average of 988.74. Only one participant reaches the top boundary of the sensor map’s range. One participant who was 58-year-old with kyphosis condition. All participants were clearly informed about the experiment objective, detailed procedures, and potential risk. The signed consents from all participants were obtained before the experiment.} We ensured that the chair had a stable communication with the PC running the receiving module and the user interface before each of the participants started his/her trial. Participants adjusted their posture so that the system could record the data with the correct label. The user interface gave the start signal and the end signal for each posture.

\begin{figure}
\centering
\subfigure[upright]{
    \includegraphics[width=0.8in]{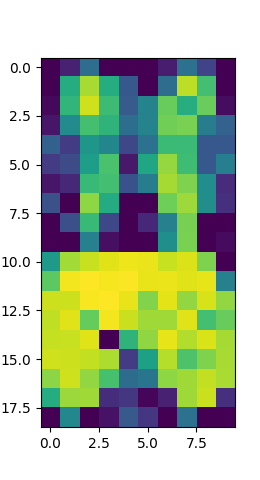}
}
\subfigure[leaning right]{
    \includegraphics[width=0.8in]{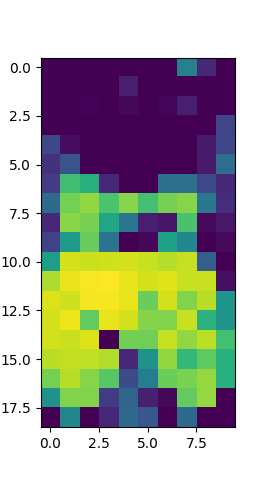}
}
\subfigure[leaning left]{
    \includegraphics[width=0.8in]{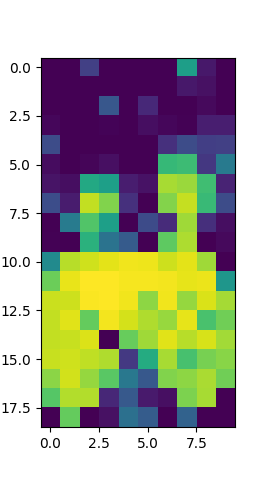}
}
\subfigure[leaning forward]{
    \includegraphics[width=0.8in]{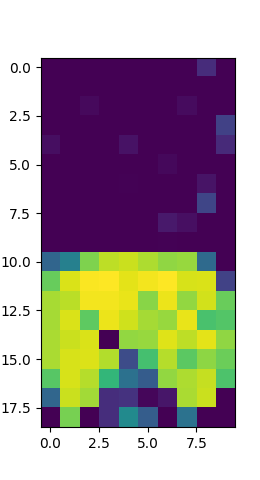}
}
\subfigure[leaning back]{
    \includegraphics[width=0.8in]{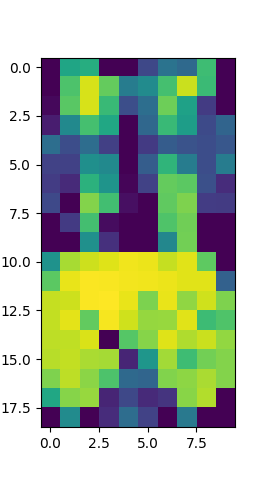}
}
\subfigure[LC and seated upright]{
    \includegraphics[width=0.8in]{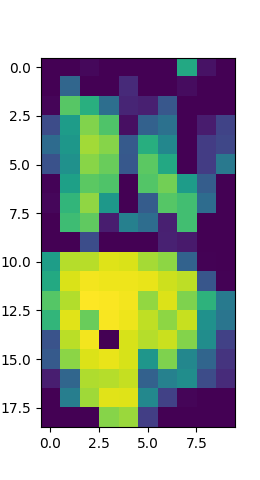}
}
\subfigure[RC and seated upright]{
    \includegraphics[width=0.8in]{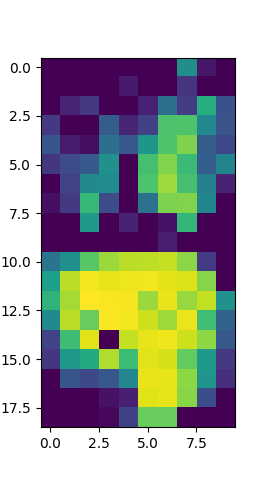}
}
\subfigure[LC and leaning back]{
    \includegraphics[width=0.8in]{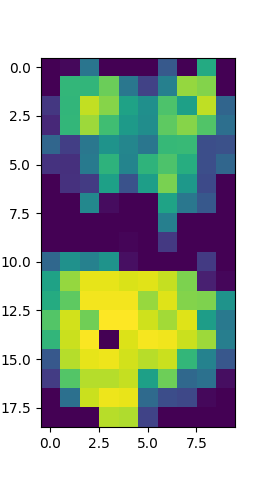}
}
\subfigure[RC and leaning back]{
    \includegraphics[width=0.8in]{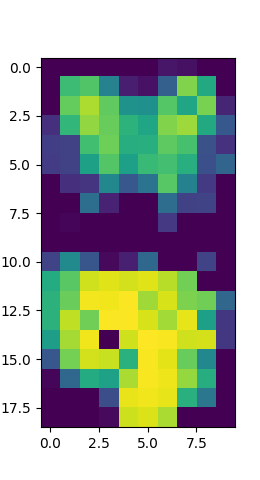}
}
\subfigure[LA and seated upright]{
    \includegraphics[width=0.8in]{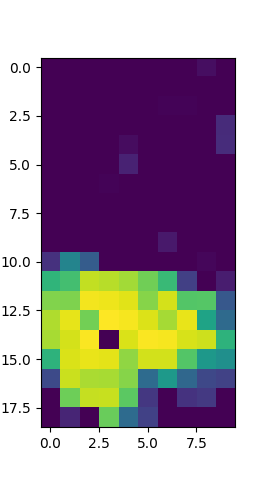}
}
\subfigure[RA and seated upright]{
    \includegraphics[width=0.8in]{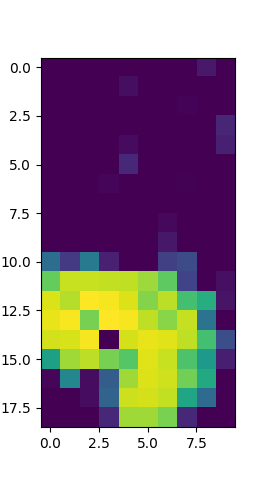}
}
\subfigure[LA and leaning back]{
    \includegraphics[width=0.8in]{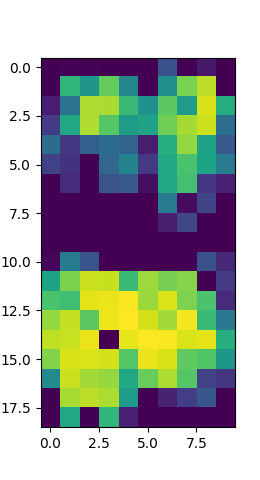}
}
\subfigure[RA and leaning back]{
    \includegraphics[width=0.8in]{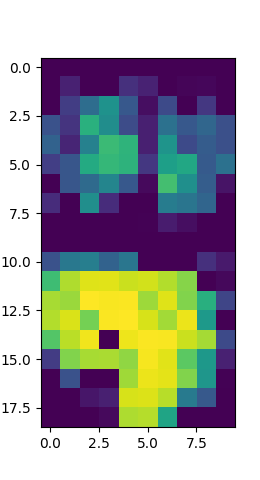}
}
\subfigure[sitting on the leading edge]{
    \includegraphics[width=0.8in]{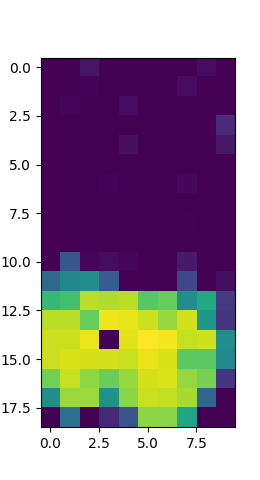}
}
\subfigure[slouching back down]{
    \includegraphics[width=0.8in]{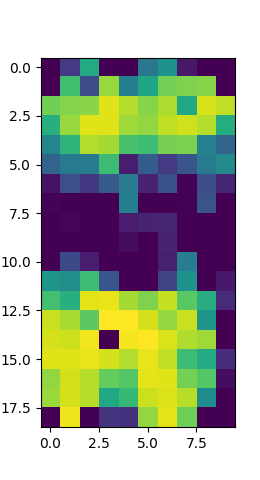}
}
\caption{View of the pressure map for the example sitting postures. All of these pressure maps are of the same participant.}
    \label{fig:data}
\end{figure}

Incorrect data were generated when participants changed their posture before the end signal or after the start signal. We cleaned the data to filter out the unreliable data and divided the rest of the data into a training set and validation set following the random shuffle principle. The data were divided into 4755 pressure maps for training and 1189 pressure maps for testing. The pressure maps of the 15 postures are shown in Fig \ref{fig:data}.

\subsection{Experiment Setup}
We set up 2 groups of models for testing the performance of our SNN. The first model was an LR classifier. The second model was an LSM with LR as its readout layer. In the first group, two trials were performed separately with encoded data and nonencoded data. In the second group, three trials performed with nonencoded data and 2 sets of encoded data with different sparsities.

In the experiment, we set the value of $C$ in equation \ref{eq:prob} to 0.3 $(EE)$, 0.2 $(EI)$, 0.4 $(IE)$, and 0.1 $(II)$.  $\lambda$ was set to 1.6667. We set the number of E neurons to 1600, the number of I neurons to 400. The number of input neurons is depending on the coding number n. For the case when n equals 1, the number of input neurons is 190. For the case n equals 2, the number of input neurons is 380. The number of output neurons was the sum of the E neurons and I neurons. The activation threshold of each neuron was 15 mV, the resting potential was 13.5 mV, the refractory period of the E neurons was 3 ms, and the refractory period of the I neurons was 2 ms. We set the amplitude of $A$ to 30 and correspondingly duration for each simulation to 30 ms. The weights were initialized with random numbers that followed a gamma distribution with a shape $\alpha$ of 30 $(EE)$, 60 $(EI)$, 19 $(IE)$, 19 $(II)$, 18 (input to E), 9 (input to I).

\subsection{Experiment Results}
The performance comparison of the above 2 models is shown in table \ref{table:result}.

The pressure map data collected by our experiment have a large interclass distance. Thus, even with only logistic regression, we obtained a satisfactory classification accuracy. In the second trial, we encoded the data. The results showed that the encoded data still can be classified by logistic regression. The third and fourth trials verified the effectiveness of the structure used in this paper. Furthermore, encoding the mapping data to the spatiotemporal sequence was necessary for the SNN. Our coding method can modify the number of coded neurons by changing the coding number $n$. In the experiments, $n=2$ achieved better results than $n=1$. However, the classification accuracy and the value of $n$ are not positively correlated. The runtime for the proposed system could reach 10 fps with an Intel(R) Core(TM) i7-9700K CPU.

\begin{table}[ht]
\caption{Performance comparison of the 4 models}
    \centering
    \begin{tabular}{cccccc}
     \toprule
     Trial& Encoding& Model& Precision& Recall& F1 score \\
     \midrule
    1& -& LR& 0.9973& 0.9971& 0.9972\\ 
    2& Our method (n=1)& LR& 0.9984& 0.9980& 0.9982\\
    3& -&  {SNN+LR}& 0.7994& 0.7773& 0.7758 \\
    4& Our method (n=1)&  {SNN+LR}& 0.9989& 0.9987& 0.9988\\
    5& Our method (n=2)&  {SNN+LR}& \textbf{0.9995}& \textbf{0.9994}& \textbf{0.9994}\\
     \bottomrule
    \end{tabular}
\label{table:result}
\end{table}
To verify the availability and the capacity of the SNN and the proposed encoding method, the data-set was split by subjects for training and testing respectively. The training set contains subject 1 to 15, and the testing set contains subject 16 to 19. The experiments parameters setting is as same as the preceding description. The results are shown in Table \ref{tab:result-2}.

\begin{table}[ht]
    \centering
        \caption{Performance comparison of the 4 models on the dataset split by subjects}
    \begin{tabular}{cccccc}
    
     \toprule
      {Trial}&  {Encoding}&  {Model}&  {Precision}&  {Recall} &  {F1 score} \\
     \midrule
      {1}&  {-}&  {LR}&  {0.8934} &  {0.8304} &  {0.8350} \\
      {2}&  {Our method (n=1)}&  {LR}&  {0.7314} &  {0.7077} &  {0.7083}\\
      {3}&  {-}&  {SNN+LR}&  {0.0061} &  {0.0714} &  {0.0112}\\
      {4}&  {Our method (n=1)}&  {SNN+LR}&  {0.8852} &  {0.8606} &  {0.8612}\\
     \bottomrule
    \end{tabular}

    \label{tab:result-2}
\end{table}
We made a performance comparison with the commonly used machine learning algorithms with our proposed approach based on our collected dataset. These mentioned algorithms include Self-Organizing Maps (SOM), Principal Component Analysis (PCA), Linear Discriminant Analysis (LDA), Locally Linear Embedding (LLE), Laplacian Eigenmaps (LE), Support Vector Machine (SVM), Random Forest (RF), K-Means and K-Nearest Neighbor (KNN). The comparison results in Table III show that our method is superior than others on classification accuracy. In these results, the results are close on the single classifier algorithm (LR, LDA, SVM, KNN and RF), in which LR has the highest accuracy. The combination of the dimension reduction algorithm and the logistic regression (PCA + LR, LLE + LR, LE + LR) are similar to our approach. This combination algorithm's performance depends on the processed data from the front-end algorithms. Different from supervised algorithms, the  clustering algorithms (SOM, K-Means) do not well perform on the sitting posture recognition task. Compared with ANN, SNN with the same structure has lower energy consumption and fewer operations \cite{rueckauer2017conversion}\cite{merolla2014million}. Thus, SNN is more suitable for embedded devices such as our sitting posture recognition system. To verify the SNN and the ANN has similar performance, we implemented a three-layer fully connection ANN and evaluated it on our dataset. The ANN is composed of a 190-neuron input layer, a 1600-neuron hidden layer, and a 15-neuron output layer. When the hidden layer adopts the 99\% of dropout as SNN, the loss hardly decreases. The precision of the ANN model on our dataset is 0.85, whereas that of SNN is 0.89. Therefore, the performance gap between ANN and SNN is not significant.

Computational complexity is one of the key problems in machine learning. Algorithms with similar performance may have different computational cost. To assess the complexity of learnable classes, \cite{blumer1989learnability} introduced the Vapnik-Chervonenkis (VC) dimension, a simple combinatorial parameter of the class of concepts. As the scores of LDA and LR in Table \ref{tab:result-3} are relatively close to the proposed method, it is appropriate to compare the computational complexity with them. Both LDA and LR are linear classifiers, which are well known to have the VC-dimension $O(N+1)$, where $N$ is the number of features. The VC-dimension of a formal model SNN can be very large, which has been proved by Wolfgang \cite{maass1995computational}. Cascaded by a linear classifier can result in the possibility of a better control of the classifier space complexity. Due to the fact that SNN part can be considered as an encoder mapping the input data to 2000-dim followed by a linear classifier, the VC-dimension of our proposed method has a lower bound $O(N+1)$. Thus, the proposed method has a better learning ability whereas is harder to converge than pure linear classifiers.
\begin{table}[ht]
    \centering
    \caption{Performance comparison with commonly used machine learning algorithms}
    \begin{tabular}{cccc}
    \toprule
          {Algorithm}&  {Precision}& {Recall}& {F1 Score} \\
         \midrule
          {LDA}& {0.86}& {0.85}& {0.84}\\
          {SVM\cite{ding2017estimation}}& {0.82}& {0.78}& {0.76}\\
          {KNN}& {0.72}& {0.68}& {0.67}\\
          {RF}& {0.85}& {0.83}& {0.83}\\
          {LR}& {0.89}& {0.83}& {0.84}\\\hline
          ANN\cite{huang2017smart}\cite{mota2003automated} & {0.85} & {0.71} & {0.76}\\\hline
          {PCA\cite{tan2001sensing}+LR}& {0.85}& {0.81}& {0.80}\\
          {LLE+LR}& {0.58}& {0.50}& {0.48}\\
          {LE+LR}& {0.01}& {0.09}& {0.02}\\\hline
          {SOM}& {0.46}& {0.42}& {0.40}\\
          {K-Means}& {0.03}& {0.06}& {0.04}\\\hline
          {Ours}& {0.89}& {0.86}& {0.86}\\
         \bottomrule
    \end{tabular}
    
    \label{tab:result-3}
\end{table}
\begin{table}[ht]
\caption{Sitting posture recognition system comparison}
\centering
\begin{tabular}{lllll}
\toprule
Algorithm & Sensor& Pos.Num\\
\midrule
ANN\cite{huang2017smart}& 8x8x1 FSR grid& 8\\
SVM\cite{ding2017estimation}& 16x16x2 FSR grid& 8\\
Determine\cite{ahn2015implemented}& 8 FSR sensors& 8\\
PCA\cite{tan2001sensing}& 42x48 Tekscan BPMS& 14\\
ANN\cite{mota2003automated}& 42x48 Tekscan BPMS& 9\\
SIR\cite{zhu2003template}& 42x48 Tekscan BPMS& 10\\
SNN+LR (ours)& 9x9 and 9x10 FSR grids& 15\\
\bottomrule
\end{tabular}
\label{table:contrast}
\end{table}

The comparison between the previously developed sitting posture recognition systems and our proposed one is shown in Table \ref{table:contrast}. It is noted that all approaches are based on the pressure distribution upon sensor grids. Our system is able to recognize the largest sitting posture classes compared with other systems.
\section{Conclusion}
\label{sec:Conclusions}
A sitting posture recognition system is presented in this work. The designed system is cost-effective with high accuracy for sitting posture monitoring. For the hardware, we used an FSR matrix as the sensor and verified that the low-density pressure matrix is sufficient to distinguish between different sitting postures. In terms of algorithms, we verified that the SNN could solve the image-like classification problem with high accuracy. Our proposed system could provide timely haptic feedback based on the accurate classification of 15 common sitting postures.



%



\ifCLASSOPTIONcaptionsoff
\newpage
\fi




\bibliographystyle{IEEEtran}
\bibliography{reference.bib}
%

%
\vspace{-12 mm}
\begin{IEEEbiography}[{\includegraphics[width=1in,height=1.25in,clip,keepaspectratio]{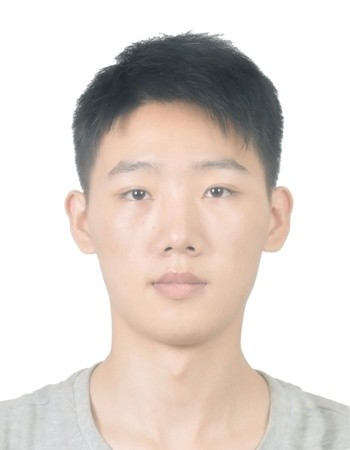}}]{Jianquan Wang}
received the B.Eng. degree in Electrical Engineering from Southwest Jiaotong University, Chengdu, China, in 2017. He is currently working toward the M.Sc. degree at the Multimedia Computing Research Laboratory, University of Ottawa, Ottawa, ON, Canada. His research interests include computer vision and deep learning.
\end{IEEEbiography}
\vspace{-0 mm}
\begin{IEEEbiography}[{\includegraphics[width=1in,height=1.25in,clip,keepaspectratio]{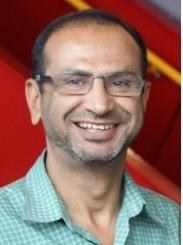}}]{Basim Hafidh}
 received his B.A.Sc in Electrical Engineering from the University of Baghdad, Baghdad, Iraq in 1981. He then received his M.A.Sc in Electrical Engineering from the same university in 1985. He received a second M.A.Sc and Ph.D. Degree in Electrical and Computer Engineering from the University of Ottawa in 2012 and 2017, respectively. He is currently a Post-Doctoral Fellow with the Multimedia Communication Research Laboratory (MCRLab), School of Electrical Engineering and Computer Science (SEECS) at the University of Ottawa. His research interests include tangible user interfaces, multi-model interaction with environment, IoT, smart homes, smart cities and haptic sensors and actuators.
\end{IEEEbiography}
\vspace{-0 mm}
\begin{IEEEbiography}[{\includegraphics[width=1in,height=1.25in,clip,keepaspectratio]{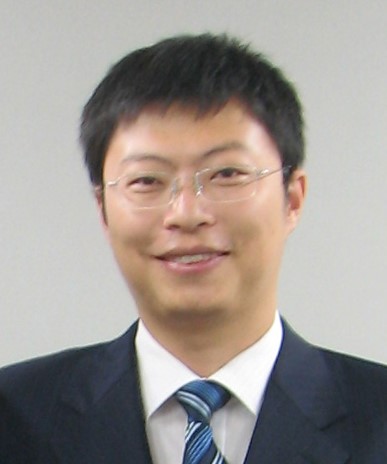}}]{Haiwei Dong}
(M'12--SM'16) received the Dr.Eng. degree in computer science and systems engineering from Kobe University, Kobe, Japan, and the M.Eng. degree in control theory and control engineering from Shanghai Jiao Tong University, Shanghai, China, in 2010 and 2008, respectively. He was a Research Scientist with the University of Ottawa, Ottawa, ON, Canada; a Postdoctoral Fellow with New York University, New York City, NY, USA; a Research Associate with the University of Toronto, Toronto, ON, Canada; a Research Fellow (PD) with the Japan Society for the Promotion of Science, Tokyo, Japan. He is currently a Principal Engineer with Huawei Technologies Canada, Ottawa, and a registered Professional Engineer in Ontario. His research interests include artificial intelligence, robotics, and multimedia.
\end{IEEEbiography}
\vspace{-12 mm}

\begin{IEEEbiography}[{\includegraphics[width=1in,height=1.25in,clip,keepaspectratio]{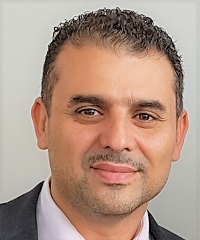}}]{Abdulmotaleb El Saddik}
(M'01--SM'04--F'09) is Distinguished University Professor and University Research Chair in the School of Electrical Engineering and Computer Science at the University of Ottawa. His research focus is on the establishment of Digital Twins to facilitates the well-being of citizens using AI, IoT, AR/VR and 5G, hence allowing people to interact in real-time with one another as well as with their smart digital representation. He has co-authored 10 books and more than 550 publications and chaired more than 50 conferences and workshop. He has received research grants and contracts totaling more than $\$$20 M. He has supervised more than 120 researchers and received several international awards, among others, are ACM Distinguished Scientist, Fellow of the Engineering Institute of Canada, Fellow of the Canadian Academy of Engineers and Fellow of IEEE, IEEE I$\&$M Technical Achievement Award, IEEE Canada C.C. Gotlieb (Computer) Medal and A.G.L. McNaughton Gold Medal for important contributions to the field of computer engineering and science.
\end{IEEEbiography}




\end{document}